\documentclass{emulateapj}
\usepackage{natbib}
\usepackage{graphicx}

\bibpunct{(}{)}{;}{a}{}{,}

\begin{document}                          

\title{Formation of Globular Clusters in Galaxy Mergers}    

\author{Yuexing Li, Mordecai-Mark Mac Low}
\affil{Department of Astronomy, Columbia University, New York,
NY 10027, USA}
\affil{Department of Astrophysics, American Museum of Natural
History, 79th Street at Central Park West, New York, NY 10024-5192, USA}
\and 
\author{Ralf S. Klessen}
\affil{Astrophysikalisches Institut Potsdam, An der Sternwarte
16, D-14482 Potsdam, Germany} 
\email{yxli@astro.columbia.edu, mordecai@amnh.org, rklessen@aip.de} 

\begin{abstract}
  We present a high-resolution simulation of globular cluster
  formation in a galaxy merger. For the first time in such a
  simulation, individual star clusters are directly identified and
  followed on their orbits.  We quantitatively compare star
  formation in the merger to that in the unperturbed galaxies.
  The merging galaxies show a strong starburst, in sharp contrast to
  their isolated progenitors. Most star clusters form in the tidal
  features. With a mass range of $5\times10^{5}$--$5\times 10^{6}
  M_{\odot}$, they are identified as globular clusters. The
  merger remnant is an elliptical galaxy. Clusters with different mass
  or age have different radial distributions in the galaxy. 
  Our results show that the high specific frequency and bimodal distribution
  of metallicity observed in elliptical galaxies are natural products of
  gas-rich mergers, supporting a merger origin for the ellipticals and
  their globular cluster systems.
\end{abstract}


\section{INTRODUCTION}
Interactions and mergers drive galactic evolution and cause starbursts.
In the past decade, the
Hubble Space Telescope (HST) has unveiled evidence for the formation
of massive star clusters or young globular clusters in merging
galaxies, such as NGC 7252 \citep{whitmore93}, the Antennae Galaxy
\citep{whitmore95}, NGC 3597 \citep{holtzman96}, and NGC 3921
\citep{schweizer96}, as reviewed by \citet{whitmore01}.
 
There have been several theoretical models for the formation of
globular clusters (GCs), including primary models where clusters form before
galaxies (e.g \citealt{peebles68}), secondary models where clusters form with
galaxies (e.g. \citealt{fall85, harris94}), tertiary models where clusters
form after galaxies, such as in mergers (\citealt{ashman92}), and unified
models \citep{elmegreen97}. Models have been reviewed by \citet{harris91,
ashman98}, and \citet{carney01}. The HST observations have prompted particular
interest in the merger scenario. There have been many numerical simulations of
galaxy mergers (\citealt{toomre72, white78, farouki82, barnes92, hernquist92,
mihos94, mihos96, dubinski99, springel99, springel00, naab01, barnes02}), but
only a few focus on GC formation in mergers. These include semi-analytical
models \citep{beasley02}, combined N-body and smoothed particle hydrodynamics
(SPH) simulations \citep{bekki02a,bekki02b}, and adaptive-grid cosmological
simulations \citep{kravtsov03}.  In these models individual GCs are not
identified and directly followed. In addition, the star formation history of
the merger is not compared to that of the isolated progenitors.

We here model GC formation in both isolated disk galaxies and their
mergers using N-body/SPH simulations. Absorbing sink particles are
used to directly represent individual massive young star clusters. In
\S~2 we briefly describe our computational method. We discuss the
burst of GC formation during the merging event in \S~3, and compare
with the evolution of the isolated galaxy. We then focus on the
cluster mass and age distribution at the end of the merger in \S~4, and
summarize in \S~5. 

\section{NUMERICAL METHOD}
We use the publicly available SPH code GADGET \citep{springel01}, and
implement absorbing sink particles that do not interact
hydrodynamically \citep{bate95} to directly measure the mass of
gravitationally collapsing gas. To represent GCs, we allow sinks to
form when the local density exceeds 1000 cm$^{-3}$
\citep{bromm02}. Our galaxy model consists of a dark matter halo, and
a disk of stars and isothermal gas. We follow the analytical work by
\citet{mo98} and the numerical implementation by \citet{springel99}
and \citet{springel00}.  Our simulations meet three numerical
criteria, the Jeans resolution criterion (\citealt{bate97}), the
gravity-hydro balance criterion for gravitational softening
\citep{bate97}, and the equipartition criterion for particle masses
\citep{steinmetz97}.  We adopt a halo concentration parameter $C = 5$,
a spin parameter $\lambda = 0.05$, and Hubble constant $H_0 = 70$ km
s$^{-1}$ Mpc$^{-1}$ \citep{springel00}.
 
The galaxy model studied here initially has rotational velocity $V_{200}$ = 
100 km s$^{-1}$ at the virial radius where the overdensity is 200, and a
virial mass of $M_{200} = 3.3\times 10^{11} M_{\odot}$.  Total mass in the
disk and in the disk gas as fraction of the virial mass are $f_{\rm d} = 0.05$ 
and $f_{\rm g} = 0.01$, respectively.  An isothermal equation of state
with sound speed $c_{\rm s} = 6$ km s$^{-1}$ is used. The total
particle number for each single galaxy is $N_{\rm tot}=1\times 10^6$,
with $N_{\rm g} = 5\times10^5$, $N_{\rm d}=2\times10^5$, and $N_{\rm
h}=3\times10^5$ for disk gas and stars, and halo dark matter, 
respectively. The corresponding gravitational softening lengths are
$\epsilon_{\rm g} = 0.01$~kpc, $\epsilon_{\rm d} = 0.1$~kpc, and
$\epsilon_{\rm h}=0.4$ kpc. The spatial resolution of the gas is thus
10 pc, and the mass per gas particle is $6.6 \times 10^3 M_{\odot}$.
This is two orders of magnitude better resolution than the models of
\citet{bekki02b}. 

We perform two simulations. One follows the evolution of the isolated
galaxy described above, and the other is an equal-mass, head-on merger
of two such galaxies. The two galaxies are on a parabolic orbit and
are separated initially by a distance of 300 kpc. In the merger run,
$N_{\rm tot} = 2\times 10^6$, with $N_{\rm g} = 1\times 10^6$ gas
particles and corresponding other particle numbers. The simulations
are followed for up to 5 Gyr.

\section{STARBURSTS IN THE MERGING EVENT}
\subsection{Where Do Globular Clusters Form?}
Our single galaxies begin marginally stable to gravitational instability
\citep{rafikov01}, with initial Toomre instability parameter for gas $Q =
1.4$. Stars form slowly but steadily, mostly in the center. In the merger
event, the galaxies first collide at $t \sim 1.4$ Gyr, then separate to $\sim
200$~kpc at $t\sim 2$ Gyr, fall back, and finally merge at $t\sim 4.5$ Gyr.  
Vigorous starbursts occur in each of the two close encounters.

\begin{figure}[h]
\begin{center}
\includegraphics[height=3.2in]{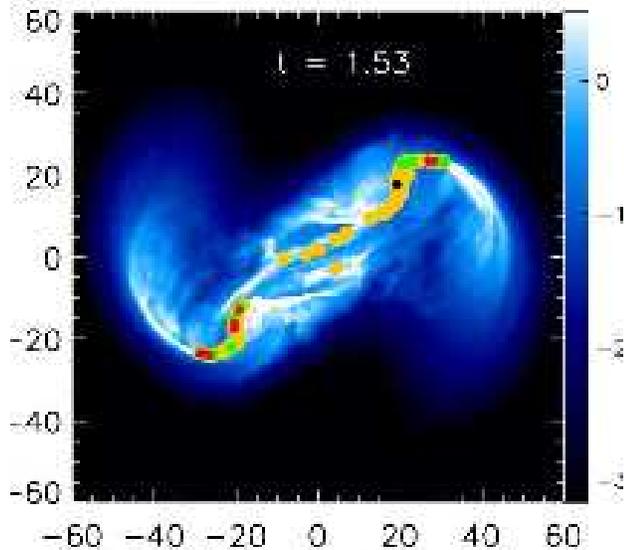}
\caption{\label{fig_tail} GC`s formed in the merger. The
blue-white image is the gas surface density at $t = 1.53$~Gyr, with values
given by the color bar.  The colored dots represent ({\em yellow}) older GCs
with lifetimes $\tau_{\rm gc} > 10$~Myr, ({\em green}) young GCs with
$\tau_{\rm gc} < 10$~Myr and $M_{\rm gc} < 10^{6} M_{\odot}$, ({\em red})
young GCs with $M_{\rm gc} \ge 10^{6} M_{\odot}$, and ({\em black}) the most
massive young GC.}  
\end{center}
\end{figure}

Most clusters form during the first close encounter at $t=$1.4--1.7~Gyr.  
To estimate their stellar masses, we make the assumption that
individual sink particles represent regions of dense molecular gas.
Observations by \citet{young99} and \citet{wong02} suggest that the {\em
local} star formation efficiency in molecular clouds remains roughly
constant. We assume that 35\% of the mass of each sink particle forms stars,
and will show  in future work that a range of 20--50\% is consistent
with the observed Schmidt law. Using this assumption, the cluster mass range
is $5\times10^{5} M_{\odot}$ to $5\times 10^{6} M_{\odot}$.  We therefore
identify these clusters as newly formed GCs.

Figure \ref{fig_tail} shows that during the first encounter GCs form
in the extended tidal features where the gas reaches high density, in
agreement with observations (\citealt{whitmore95, zhang01}). The
derived mass range agrees well with spectroscopic estimates by
\citet{mengel02} of some young clusters in the Antennae Galaxy.

\subsection{Comparison Between Merger and Isolated Galaxy} 
\begin{figure}[h]
\begin{center}
\includegraphics[height=3.2in]{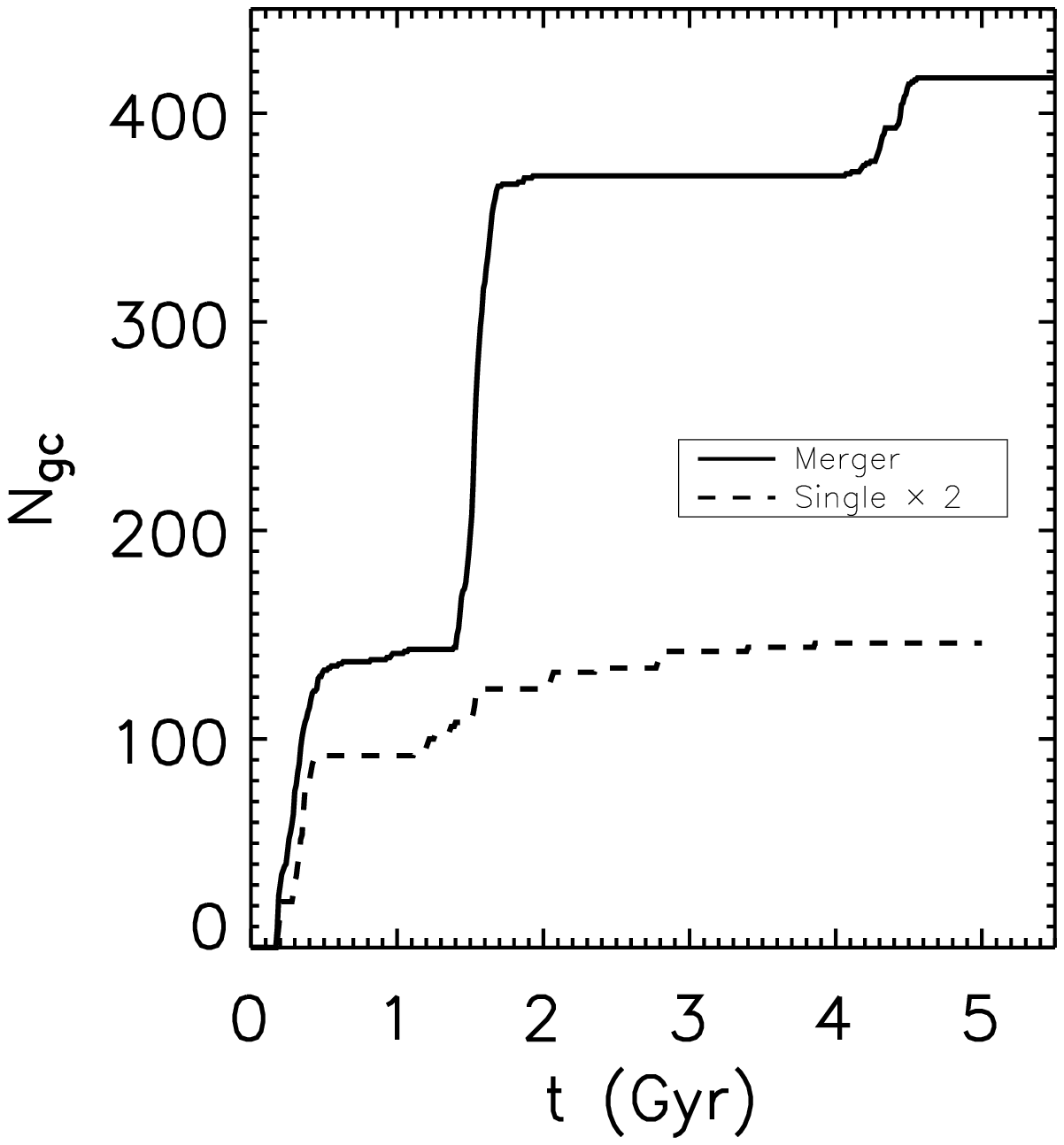}\\
\includegraphics[height=3.2in]{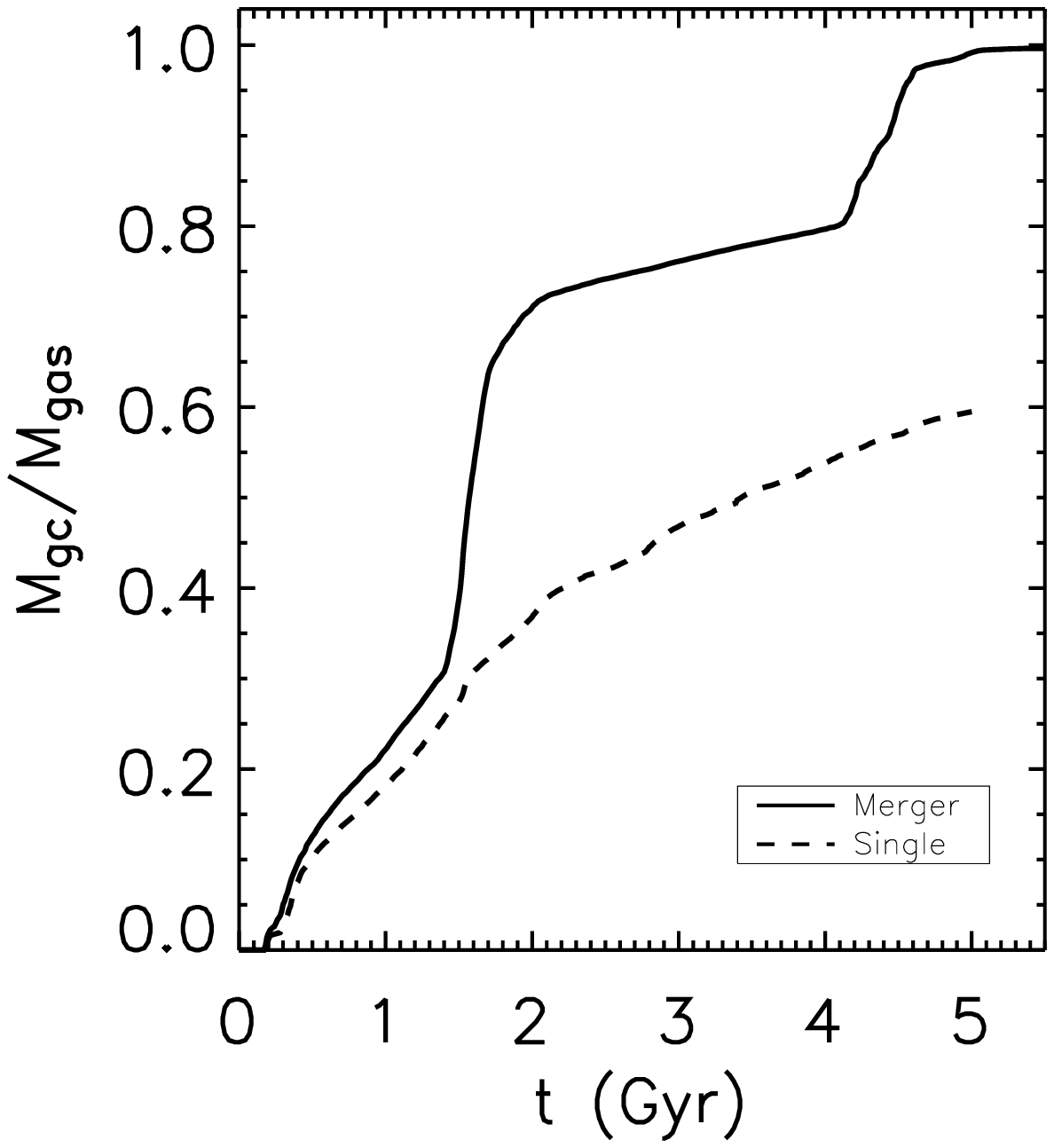}
\caption{\label{fig_com} Number $N_{\rm gc}$ (\textit{top}) and
  fractional mass $M_{\rm gc}/M_{\rm gas}$ (\textit{bottom}) of GCs
  over time compared between two merging galaxies ({\em solid lines})
  and the same two galaxies isolated ({\em dashed lines}). Note the sudden
  jumps in the merging galaxies, representing starbursts. }
\end{center}
\end{figure}

Figure
\ref{fig_com} compares star formation in the isolated galaxy with the galaxy
merger. In the isolated galaxy, stars start to form at $t \sim 0.3$~Gyr when
gravitational instability increases the gas density. The cluster number
$N_{\rm gc}$ and mass $M_{\rm gc}$ slowly increase as the galaxy maintains
marginal stability. During the merging event, the gas density is locally
increased  by shocks and tidal forces, leading to starburst behavior. The
cluster formation rate $dM_{\rm gc}/dt$ increases dramatically, particularly
in the first encounter when more gas is available. At the end of the
simulation, $N_{\rm gc} = 73$ for the isolated galaxy and 417 for the merger.

The specific frequency \citep{harris81} of GCs in a galaxy is $S_{\rm N} =
N_{\rm t}10^{0.4({\cal M}_{\rm v} + 15)}$, where $N_{\rm t}$ is the total
number of GCs, and the V-band absolute magnitude of the host galaxy, ${\cal
M}_{\rm v} = {\cal M}_{\odot} -2.5$log($L_{\rm v}/L_{\odot}$). We use the
method of \citet{bekki02b} to estimate $S_{\rm N}$ in our models by assuming
$M_*/L_v \sim 3$, as in the Milky Way. In order to account for the cluster
destruction mechanisms described by \citet{spitzer87,gnedin97}, and
\citet{fall01}, we assume a 10\% survival rate of our clusters. Our final
derived value of $S_{\rm N}$ depends linearly on this. Each single galaxy has
a stellar mass of $M_* \sim 1.65 \times 10^{9} M_{\odot}$, so we can derive
${\cal M}_{\rm v} \sim -17.1$~mag, so $S_{\rm N} \sim 1.0$. Similarly, the
elliptical merger remnant has $M_* \sim 3.3 \times 10^{9} M_{\odot}$, which
gives ${\cal M}_{\rm v} \sim -17.9$~mag, and $S_{\rm N} \sim 3.0$. These
values are consistent with observations.  For example, the Milky Way-like
spiral M31 has $S_{\rm N} \sim 1.2 \pm 0.2$ \citep{barmby01}, while typical 
ellipticals have $S_{\rm N} \sim 3.5$ (\citealt{harris91, mclaughlin99}).
These estimates do depend on the stellar mass-to-light ratio of the galaxies
(e.g.\ \citealt{larsen02, rz04}) and survival rate of the clusters
(\citealt{spitzer87, gnedin97, fall01, whitmore04}). Nevertheless, they
reflect the difference between isolated disk galaxies and  mergers, and
suggest that mergers can explain the considerably higher specific frequency of 
GCs observed in elliptical galaxies.  

\section{MERGER REMNANT}
\begin{figure}[h]
\begin{center}
\includegraphics[height=3.2in]{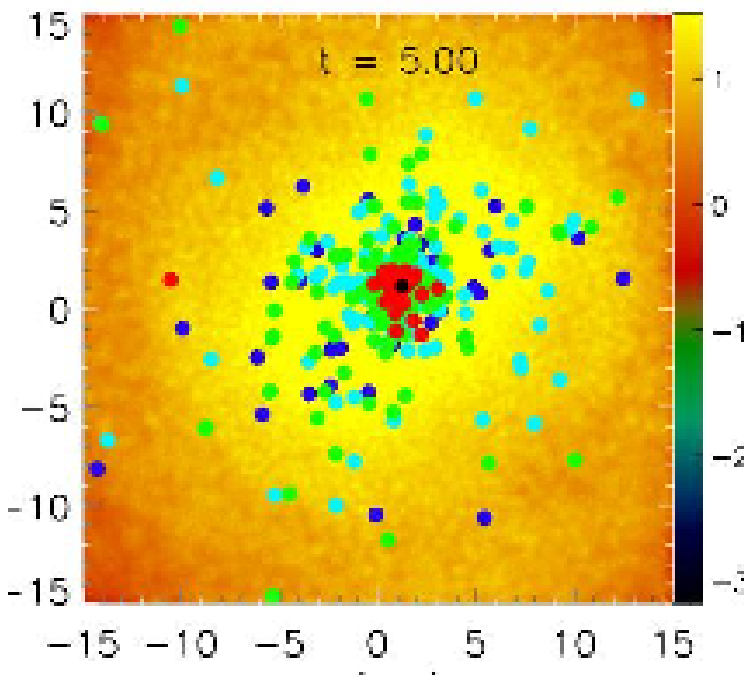}\\
\includegraphics[height=3.2in]{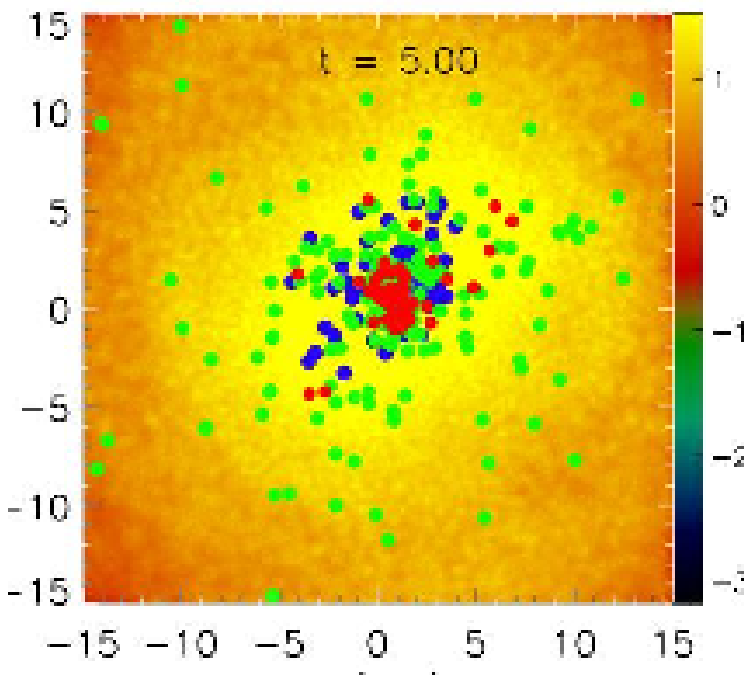}
\caption{\label{fig_dist} Log of stellar surface density in the merger
remnant, with values shown in color bar, compared to distribution of GCs ({\em
dots}) in mass (\textit{top panel}) and age ({\em bottom panel}). Mass bins
shown are $M_{\rm gc} < 10^{6} M_{\odot}$ ({\em blue}), $ 10^{6} \le M_{\rm
gc} < 10^{6.5} M_{\odot}$ ({\em cyan}), $10^{6.5} \le M_{\rm gc} < 10^{7}
M_{\odot}$ ({\em green}), $M_{\rm gc} \ge 10^{7} M_{\odot}$ ({\em red}), and
the most massive GC, $M_{\rm gc} = 7.8\times 10^7 M_{\odot}$ ({\em
black}). Age bins shown are GCs formed at $t < 1.4$~Gyr during initial
approach before the merging event begins ({\em blue}), GCs formed at $1.4$ Gyr
$ < t < 1.9$ Gyr during the first encounter ({\em green}), and GCs formed at
$4.0$ Gyr $ < t < 4.6$ Gyr during the second encounter ({\em red}).}
\end{center}
\end{figure}

Star formation during the merger leads to rapid gas depletion.  By
$t=5$~Gyr the two merging galaxies have transformed into an elliptical
galaxy, whose stellar distribution is shown in Figure \ref{fig_dist}.
Three distinct populations of GCs form at different epochs, those
formed during the two close encounters are almost coeval, as they form
over a very short period during the encounters.

Figure~\ref{fig_dist} shows the GC system is centered on the stellar
spheroid. Massive GCs are concentrated towards the center of the elliptical
galaxy that remains after the merger, while less massive ones are
found at greater radii as well.  Old GCs formed prior to the first encounter
and the youngest GCs formed in the second encounter are 
concentrated towards the center of the galaxy, while GCs formed in the
first encounter are also found at large radii.  The oldest GCs in our
simulation are also the most massive, and quickly sink towards the
center by dynamical friction. The youngest ones occur in the galactic center
because remaining gas assembles there during the final merger. During the
first encounter, on the other hand, GCs form in the tidal tails with
greater angular momentum, and so have a broad radial distribution.

If we assume a metallicity-age relation such that older GCs have lower
metallicities \citep{rich01}, we clearly see a bimodal metallicity
distribution. Young, high-metallicity GCs are centrally concentrated,
while older, lower-metallicity GCs extend to larger radii, in
agreement with observations \citep{djorgovski94}. The isolated galaxy forms
clusters slowly and steadily, so age, as well as metallicity, follow a smooth
distribution. There is no bimodality.  Our results thus support the arguments
of \citet{ashman92} and \citet{kundu01} that mergers are required to produce a
bimodal distribution. 

\section{SUMMARY}
We present high-resolution simulations of star cluster formation in
both a single galaxy and a major merger, using a three-dimensional SPH
code that includes absorbing sink particles to represent massive star
clusters. This allows direct identification of individual clusters and
tracking of their orbital evolution over several gigayears. The
merging galaxies show bursts of massive star cluster formation, in
sharp contrast to the steady but slow formation in an isolated galaxy.
Most new clusters form in the tidal tails and bridges between the
merging galaxies. They are identified as progenitors of globular
clusters, although it should be emphasized that we do not include
dynamical destruction of clusters (e.g. \citealt{spitzer87,gnedin97,fall01})
in the simulations. Dynamical destruction will change the number and mass of
GCs, affecting our derived specific frequency (where we accounted for it
very crudely) and spatial distribution.  Inclusion of a destruction
model to determine the evolution of our cluster particles will be
necessary to fully address these questions.

The estimated specific globular cluster frequency $S_{\rm N}$ in the
elliptical galaxy resulting from the merger exceeds by a factor of
3 that in an isolated galaxy with the same parameters as the
merging galaxies, most of the enhancement being in metal-rich GCs. 
This supports the idea that the higher $S_{\rm N}$ of metal-rich GCs observed 
in ellipticals is produced by mergers. However, elliptical galaxies also show
higher $S_{\rm N}$ for metal-poor GCs. Early mergers could explain this,
though other mechanisms have also been proposed \citep{rz04}. Clusters formed
during different phases of the merger have distinct radial distributions. The
spatial distribution of metal-poor old clusters formed before the first 
encounter as well as of metal-rich younger clusters from the
final encounter is centrally concentrated, while clusters formed
in the first encounter are more widely dispersed.  This suggests the
observed multimodal metallicity distribution of globular clusters in
elliptical galaxies is the direct result of merger processes.

\acknowledgments We thank V.\ Springel for use of both GADGET and his
galaxy initial condition generator, as well as for useful discussions.
We also thank A.-K.\ Jappsen for participating in the implementation
of sink particles in GADGET, and K.\ Bekki, O.\ Gnedin, R.\ McCray, M.
Shara, and S. Zepf for stimulating discussions, and the anonymous referee for
valuable comments that have helped to improve this manuscript. 
This work was supported by NSF grants AST99-85392 and AST03-07854, NASA grant
NAG5-13028, and DFG grant KL1358/1.  Computations were performed at the
Pittsburgh Supercomputer Center supported by the NSF, on the Parallel
Computing Facility of the AMNH, and on an Ultrasparc III cluster generously 
donated by Sun Microsystems.

\end{document}